\documentclass{elsart}

\usepackage{natbib}

 \usepackage{graphicx}
\usepackage{amssymb}

\journal{New Astronomy}

\begin{document}

\begin{frontmatter}



\title{On The Recently Discovered Pulsations From RX J1856.5-3754}


\author[Tbilisi1]{N. Chkheidze\corauthref{cor1}}
\ead{n.chkheidze@gmail.com}
 \corauth[cor1]{Corresponding author.}
\author[Tbilisi1]{D. Lomiashvili}
\ead{lomiashvili@gmail.com}

\address[Tbilisi1]{Tbilisi state university, 3 Chavchavadze Ave., 0128,
              Tbilisi, Georgia}

\begin{abstract}
An explanation of the recently discovered $7$ s pulsations from the
isolated neutron star RX J1856.5-3754 is presented. It is assumed
that the real spin period of this source is $\approx1$ s, whereas
the observed spin-modulation is caused by the presence of a nearly
transverse, very low frequency drift waves in the pulsar
magnetosphere. It is supposed that the period of the drift wave is
equal to a recently observed one. The simulated lightcurve is
plotted, the angular parameters are defined and the value of the
pulsed fraction of only $\sim 1.2\%$ is explained.
\end{abstract}


\end{frontmatter}

% main text
\section{Introduction}
\label{1} None of the previous analysis of the X-ray data of RX
J1856.5-3754 (henceforward RXJ1856) revealed any significant
periodicity \citep{pons02,ran02,drake02,burw03}. On the contrary a
recent XMM-Newton observation of RXJ1856 has discovered that this
isolated neutron star pulsates with a period of $7.055$ s
\citep{tien07}.

This object has a completely featureless X-ray and optical spectra.
The lack of any significant spectral features in the X-ray spectrum
argues against a heavy element atmosphere \citep{burw01,burw03},
whereas single temperature hydrogen atmosphere fits over-predict the
optical flux by a large factor \citep{pavlov96,pons02,burw03}. As
the soft X-ray spectrum is well fitted by the Planckian spectrum
with a temperature $63\pm3$eV \citep{burw03}. It has been proposed
that the star has no atmosphere, but a condensed matter surface
\citep{burw01,turolla04}. Such a surface might emit a featureless,
likely a blackbody spectrum, at a temperature close to that of the
surface, as suggested by \citet{pavlov00}. The overall spectra of
this source has often been described by a two-temperature blackbody
models \citep{pons02,pavlov02,burw03}, because the parameters
derived from X-rays do not fit the optical data, which shows the
Rayleigh-Jeans slope with an intensity a factor of 6 larger than
that of a X-ray emission. Alternatively a recent paper by
\citet{ho07} explains the observed featureless spectra of this
object assuming that the star has a thin magnetic, partially ionized
hydrogen atmosphere on top of a condensed surface. Though, one of
important uncertainties for this model appears to be a creation of
thin hydrogen atmospheres. However, to make these models work, the
NS has to have a condensed matter surface, which requires the
specific conditions \citep{lai97,lai01}. Also, it remains to be seen
weather a detailed analysis of magnetically condensed matter
confirms the required non-uniform distribution of the surface
temperature (two-component blackbody model). So we must conclude
that existing models, based on an assumption that the emission of
RXJ1856 has a thermal nature face numerous problems.

In the present paper we propose our explanation of the detected
pulsations in the X-ray emission of RXJ1856, which is naturally
possible based on emission model developed by \citet{ch07}. In this
paper it is assumed that the emission of this object is generated by
the synchrotron radiation, which is created as the result of
appearance of pitch angles during the quasi-linear stage of the
cyclotron instability. We suppose that the model proposed by
\citet{ch07} works, which assumes the case of a nearly aligned
rotator, whereas the periodic variations of the observed emission
may be caused by the presence of a very low frequency, nearly
transverse drift waves in the pulsar magnetosphere. These waves
propagate across the magnetic field and encircle the open field line
region of the pulsar magnetosphere \citep{kaz91b,kaz96}. They are
not directly observable but only cause the periodic change of the
direction of the pulsar emission \citep{lom06}.

In this paper, we give a description of the emission mechanism in
Section 2. The mechanism of change of the pulsar radiation direction
is described in Section 3, our model is presented in Section 4 and
conclusions are done in Section 5.

 \section{Emission mechanism}
 \label{2}

As it is known the pulsar magnetosphere is filled by a dense
relativistic electron-positron plasma. The (e$^{+}$e$^{-}$) pairs
are generated as a consequence of the avalanche process (first
described by \citet{sturrock71}) and flow along the open magnetic
field lines. The plasma is multi-component, with a one-dimensional
distribution function ( see Fig.1 from  \citet{arons81}) and
consists of the following components: the bulk of plasma with an
average Lorentz-factor $\gamma_{p}\simeq10$; a tail on the
distribution function with $\gamma_{t}\simeq10^{4}$ and the primary
beam with $\gamma_{b}\simeq10^{6}$. The main mechanism of wave
generation in plasmas of the pulsar magnetosphere is the cyclotron
instability. Generation of waves is possible if the condition of the
cyclotron resonance if fulfilled \citep{kaz91b}:
\begin{equation}
    \omega-k_{\varphi}V_{\varphi}-k_{x}u_{x}+\frac{\omega_{B}}{\gamma_{r}}=0,
\end{equation}
where $V_{\varphi}$ is the particle velocity along the magnetic
field, $\gamma_{r}$ is the Lorentz-factor for the resonant particles
and $u_{x}=cV_{\varphi}\gamma_{r}/\rho\omega_{B}$ is the drift
velocity of the particles due to curvature of the field lines
($\rho$ is the radius of curvature of the field lines and
$\omega_{B}=eB/mc$ is the cyclotron frequency). Here cylindrical
coordinate system is chosen, with the $x$-axis directed transversely
to the plane of field line, when $r$ and $\varphi$ are the radial
and azimuthal coordinates. During the quasi-linear stage of the
instability a diffusion of particles arises as along, also across
the magnetic field lines. Therefore, plasma particles acquire
transverse momenta and begin to rotate along the Larmor orbits. The
synchrotron emission is generated as a result of the appearance of
pitch angles.

In \citet{ch07} it has been assumed that the emission of RXJ1856 is
generated by the synchrotron mechanism, which is created as the
result of the cyclotron instability. Though, the original waves
excited during the cyclotron resonance come in the radio domain, the
radio emission is not observed from RXJ1856. This waves as well as
the X-ray and optical emission propagate along the local magnetic
field lines. One of the possible explanations why the radio emission
is not detected from this object is that it is generated at lower
altitudes in contrast to the X-ray and optical emission and might
miss our line of sight. Another explanation is that the radio
emission covers a large distance in the pulsar magnetosphere (since
the model of the aligned rotator is used). So there is a high
probability for it, to come in the cyclotron damping range
$\omega-k_{\varphi}V_{\varphi}-k_{x}u_{x}-\omega_{B}/\gamma_{r}=0$
\citep{khe97}. In this case the radio emission will not reach an
observer.

\begin{figure}
\centering
\includegraphics[width=5 cm]{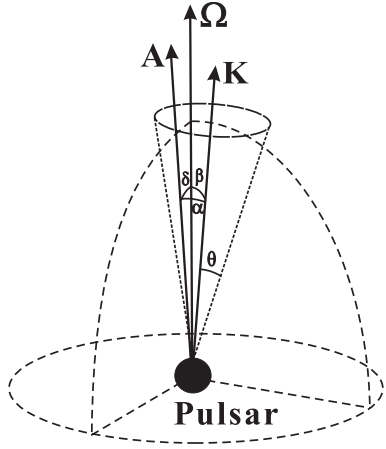} \caption{Geometry of {\bf $\Omega$} -
rotation, {\bf K} - emission and {\bf A} - observers axes. Angles
$\delta$ and $\vartheta $ are constants, while $\beta$ and $\alpha$
are oscillating with time.}
       \label{Fig1}
\end{figure}

\section{Change of the field line curvature and the emission direction by the drift waves}
 \label{3}

Considered distribution function should generate various wave-modes
in certain conditions. Particularly it has been shown
\citep{kaz91b,kaz96} that a very low frequency, nearly transverse
drift waves can be excited. They propagate across the magnetic
field, so that the angle between ${\bf k}$ and ${\bf B}$
is close to $\pi /2$. In other words, $k_{\bot }/k_{\varphi }\gg 1$, where $%
k_{\bot }=(k_{r}^{2}+k_{\varphi }^{2})^{1/2}$ . The period of the
drift waves $P_{dr}$ can be written as \citep{lom06}:
\begin{equation}
P_{dr}=\frac{e}{4\pi ^{2}mc}\frac{BP^{2}}{\gamma},
\end{equation}
where $P$ is the pulsar spin period, $\gamma$ is the Lorentz-factor
of the relativistic particles and $B=B_{s}(R_{0}/R)^{3}$ is the
magnetic field in the wave excitement region ($B_{s}$ is the
magnetic field at the pulsar surface and $R_{0}$ is the radius of
the neutron star). It appears that the period of the drift waves can
vary in a broad range.

The magnetic field of drift wave adds with the pulsar magnetic field
as $r$ component and causes changing of the curvature of field line.
Even a small change of $B_{r}$ causes significant change of $\rho $.
Variation of the field line curvature can be estimated as:
\begin{equation}
\frac{\Delta \rho }{\rho }\approx k_{\varphi }r\frac{\Delta B_{r}}{
B_{\varphi }},
\end{equation}
here $k_{\varphi }$ is a longitudinal component of wave vector and
$r$ is distance to the center of the pulsar. It follows that even
the drift wave with a modest amplitude $B_{r}\sim \Delta B_{r}\sim
0.01B_{\varphi }$ alters the field line curvature substantially
$\Delta \rho /\rho \sim 0.1$.

Since the pulsar emission propagates along the local magnetic field
lines, curvature variation causes change of the emission direction,
with the period of the drift waves.

\section{The model}
 \label{4}

There is unequivocal correspondence between the observable intensity and $%
\alpha $ (the angle between the line of sight of an observer and the
emission direction, see Fig.~\ref{Fig1}). Maximum of intensity
corresponds to the minimum of $\alpha $. The period of pulsar is the
time interval between neighboring maxima of observable intensity
i.e. minima of $\alpha $. According to this fact, we can say that
the observable period depends on time behavior of $\alpha $ and as
it appears below it might differ from the 'real' spin period of the
pulsar.

From pulsar geometry it follows that $\alpha$ can be expressed as
\citep{lom06}:
\begin{eqnarray}
\alpha =\arccos [ \sin \delta \sin ( \beta _{0}+ \Delta \beta \sin (
\omega_{dr}t+\varphi ) ) \cos \Omega t\nonumber
 \\
+ \cos \delta \cos ( \beta _{0}+ \Delta \beta \sin (
\omega_{dr}t+\varphi ) ) ],
\end{eqnarray}
where $\Omega =2\pi/P$ is the angular velocity of the pulsar, $%
\delta $ is the angle between the rotation and the observer's axes,
$\beta $ is the angle between the rotation and emission axes (see
Fig.~\ref{Fig1}) and $\Delta \beta$ is the amplitude of changing of
$\beta$.

In the absence of the drift wave $\beta =\beta _{0}=const$ and
consequently the period of $\alpha $ equals to $ 2\pi/\Omega $. On
the other hand, if the angle between the rotation and emission axes
is too small i.e. $\delta << 1$, then the period of $\alpha$ equals
to $P_{dr}=2\pi/\omega_{dr}$. In this case the observable period
$P_{obs}$ does not represent the real spin period of the pulsar, but
equals to the period of the drift wave, which we suppose to be 7.055
s. When the real spin period of this object has been estimated by
\citet{ch07} to be $P\approx1 s$.

\begin{figure}
\includegraphics[width=11 cm]{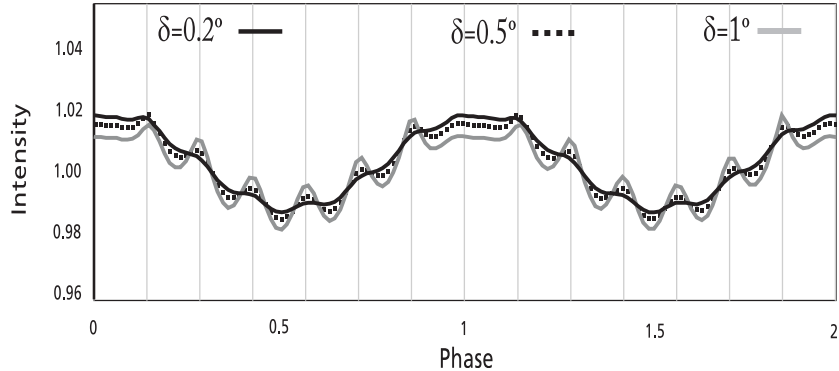} \caption{Simulated lightcurve of RX J1856.5-3754}
\label{Fig2}
\end{figure}

Hence, for some values of parameters $\beta $, $%
\Delta \beta $, $\delta $, $\varphi $ and $\vartheta $ it is
possible to explain the 7 s pulsations of RXJ1856. If we consider
this object in the framework of our model its angular parameters
will get the values shown in Table 1. Simulated lightcurve for
RXJ1856 is presented on Fig.~\ref{Fig2}, which well expresses the
value of the pulsed fraction of only $\sim 1.2\%$ (the smallest ever
seen in the isolated X-ray pulsars), obtained from observations
\citep{tien07}.

\section{Conclusion}
 \label{5}
In the present paper we propose our explanation of recently
discovered 7 s pulsations in the isolated neutron star RXJ1856,
based on model drawn by \citet{lom06}. The main feature of this work
is that the spin period of pulsar might differ from the observable
one, which is the consequence of existence of very low frequency
drift waves in the region of excitement of the pulsar emission.
These particular waves are not detected but only result in a
periodical change of curvature of the magnetic field lines, which in
turn cause the change of observed radiation with a period of the
drift wave.

Drift wave driven model is very convenient, since it makes able to
explain almost every feature of known extraordinary pulsars such as:
extremely long period radio pulsars \citep{lom06}, Rotation Radio
Transients \citep{lom07}, Anomalous X-ray Pulsars and Soft Gamma-ray
Repeaters \citep{mal07}.

We suppose that treatment of the 'real' period of RXJ1856 can be
achieved by more detailed observations. In the case of nonzero
$\delta$, which is most likely, variations with different time-scale
should be appeared with value of the pulsar spin period, see
Fig.~\ref{Fig2}. If this is confirmed, it will benefit to our model.

\section*{Acknowledgments}

We are grateful to George Machabeli for valuable discussions. This
work was partially supported by Georgian NSF Grant ST06/4-096.

\begin{table}
\caption{The values for parameters of RX J1856.5-3754}             % title of Table
\label{table:1}      % is used to refer this table in the text
\centering                          % used for centering table
\begin{tabular}{c c c c c c}        % centered columns (4 columns)
\hline\hline                 % inserts double horizontal lines
$P_{obs}(s)$ & $P(s)$ & $\Delta \beta$ & $\beta _{0}$ & $\delta$ & $\vartheta$  \\    % table heading
\hline                        % inserts single horizontal line
7.055 & $\approx 1$ & 0.02 & 0.02 & 0.01 & 0.03 \\      % inserting body of the table
\hline                                   %inserts single line
\end{tabular}
\end{table}

% Bibliographic references with the natbib package:
% Parenthetical: \citep{Bai92} produces (Bailyn 1992).
% Textual: \citet{Bai95} produces Bailyn et al. (1995).
% An affix and part of a reference:
%   \citep[e.g.][Ch. 2]{Bar76}
%   produces (e.g. Barnes et al. 1976, Ch. 2).

\end{document}